# The potential profile at the LaAlO$_3$/SrTiO$_3$ (001) heterointerface in operando conditions

M. Minohara[1,a)], Y. Hikita[1], C. Bell[1,b)], H. Inoue[2], M. Hosoda[1,3], H. K. Sato[1,3], H. Kumigashira[4], M. Oshima[5], E. Ikenaga[6], and H. Y. Hwang[1,2]

[1]*Stanford Institute for Materials and Energy Sciences, SLAC National Accelerator Laboratory, Menlo Park, California 94025, USA*

[2]*Geballe Laboratory for Advanced Materials, Department of Applied Physics, Stanford University, Stanford, California 94305, USA*

[3]*Department of Advanced Materials Science, The University of Tokyo, Kashiwa, Chiba 277-8561, Japan*

[4]*Photon Factory, Institute of Materials Structure Science (IMSS), High Energy Accelerator Research Organization (KEK), Tsukuba, Ibaraki 305-0801, Japan*

[5]*Department of Applied Chemistry, The University of Tokyo, Bunkyo-ku, Tokyo 113-8656, Japan*

[6]*Japan Synchrotron Radiation Research Institute (JASRI), SPring-8, Sayo, Hyogo 679-5198, Japan*

**Abstract**

We report measurements of the gate-bias dependent band alignment, especially the confining potential profile, at the conducting $LaAlO_3/SrTiO_3$ (001) heterointerface using soft and hard x-ray photoemission spectroscopy. Depth-profiling analysis reveals that a significant potential drop on the $SrTiO_3$ side of the interface occurs within ~2 nm of the interface under negative gate bias voltage. These results demonstrate gate control of the collapse of permittivity at the interface, and explain the dramatic loss of electron mobility with back-gate depletion.

a) Author to whom correspondence should be addressed; Electronic mail: minohara@post.kek.jp

b) Present address: H. H. Wills Physics Laboratory, University of Bristol, Tyndall Avenue, Bristol, BS8 1TL, UK

Since the discovery of a variety of interfacial electronic states between insulating and non-magnetic (001)-oriented LaAlO$_3$ and SrTiO$_3$, such as high-mobility metallic states [1], superconductivity [2,3], and magnetism [4-6], the origin of these properties has been widely discussed [7]. Given the robust insulating character of LaAlO$_3$, it is generally understood that the electron gas forms on the SrTiO$_3$ side of the interface [8-18]. Indeed, an *in situ* photoemission spectroscopy (PES) study revealed downward band bending toward the interface in the SrTiO$_3$ [15]. Notable in this system is the dramatic tunability of the interfacial conductivity using external electric fields, attracting considerable attention for fundamental studies, as well as device applications [19,20]. The application of a back-gate voltage $V_g$ tunes multiple parameters in the system simultaneously, including the superconducting transition temperature, the carrier density, the Hall mobility and the confining electric field [3,19,21-24]. In order to understand how these changes are inter-related, especially the dramatic loss of Hall mobility with back-gate depletion [22], knowledge of the band alignment and potential profile changes with gating is essential.

In this study we analyze the depth profile of the potential on the SrTiO$_3$ side of LaAlO$_3$/SrTiO$_3$ (001) heterojunctions using synchrotron radiation PES for various $V_g$. Depth resolution was achieved by varying the energy of the synchrotron-radiation light source - both soft x-ray PES (SX-PES) and hard x-ray PES (HAX-PES) were utilized - combined with precise tuning of the incident and emission angles [25,26]. Analysis of the SX-PES and HAX-PES core-level spectra with negative $V_g$ reveal a downward shift of the LaAlO$_3$/SrTiO$_3$ interface band offset, and the existence of a ~2 nm thick accumulation layer with an abrupt modulation of the electrostatic potential on the SrTiO$_3$ side of the interface. These results explain why back-gate depletion modulates the mobility far more strongly than the carrier density, and suggests this is a generic feature of nonlinear dielectrics that can be utilized in device structures.

Two LaAlO$_3$/SrTiO$_3$ heterostructures were fabricated on TiO$_2$-terminated SrTiO$_3$ (001) substrates by pulsed laser deposition, with LaAlO$_3$ thicknesses 10 and 5 unit cells (~4 and ~2 nm), as described elsewhere [22]. SX-PES and HAX-PES synchrotron radiation measurements were carried out under applied $V_g$ at beamline BL2C of the Photon Factory, KEK, Japan and beamline BL47XU of SPring-8, Japan, respectively. The SX-PES and HAX-PES spectra were recorded using a Scienta SES-2002 electron energy analyzer, and a Scienta R-4000 electron energy analyzer, respectively. A schematic of the experimental setup is shown in Fig. 1. $V_g$ was applied from the back of the 0.5 mm thick SrTiO$_3$ substrate during the PES measurements, with the LaAlO$_3$/SrTiO$_3$ interface grounded using Al wire bonding.

The measured Ti 2$p$ core-level spectra of the 4 nm and 2 nm LaAlO$_3$/SrTiO$_3$ heterostructures at $V_g$ = 0 V and a bare SrTiO$_3$ (001) substrate are shown in Fig. 2. The characteristic probing depth $\lambda$, based on the theoretical study of Tanuma *et al*. [27], was tuned by changing the irradiation energy $h\nu$ and the photoelectron emission angle $\theta$ with respect to the surface normal. $h\nu$ was 1.2 keV and 7.9 keV for the 4 nm sample and 800 eV for the 2 nm sample, $\theta$ was varied from 0 to 80 degrees for both samples (also see Fig. 1). The Ti 2*p* core-level spectra from the buried SrTiO$_3$ are shifted towards higher binding energy, as previously observed. Focusing on the 4 nm sample, as the probing depth becomes shallower ($\lambda$ < 10 nm), the peak position shifts by less than 50 meV towards higher binding energy, which is comparable to the experimental error. These results imply that the potential profile gradually varies on the scale of 10 nm in the SrTiO$_3$ from the LaAlO$_3$/SrTiO$_3$ heterointerface.

Figure 3 show the measured Ti 2*p* core-levels for the 4 nm sample using (a) SX-PES and (b) HAX-PES with -100 V ≤ $V_g$ ≤ 100 V. For all $V_g$, the gate leakage current through the SrTiO$_3$ during the PES measurements was less than 10 nA. The binding energy of SrTiO$_3$ core-level spectra were normalized to those of the LaAlO$_3$ spectra, which minimizes possible artifacts from the PES measurements, such as fluctuations of the photon energy. For negative $V_g$ the Ti 2*p* core-level first shifts to *higher* binding energy before saturating in the case of SX-PES, while it remains unchanged for the HAX-PES. Both SX-PES and HAX-PES measurements show no Ti 2*p* core-level shift for positive $V_g$. Similar results were also obtained in the Sr 3*d* core-level measurements (not shown). The change of the Ti 2*p* core-level peak position are plotted as a function of $V_g$ in Figs. 3(c) and (d), showing a total energy shift ~0.15 eV for $V_g$ < 0 V for the SX-PES.

Qualitatively we can examine these results by considering the simultaneous tuning of the sheet carrier density and the potential profile due to the application of negative $V_g$.

Considering the SrTiO$_3$ substrate as a capacitor dielectric between the back gate contact and the interface conducting layer, $V_g$ < 0 V corresponds to depletion of carriers at the interface resulting in a downward shift of the Fermi energy ($E_F$) towards the conduction band bottom. At the same time however the band-bending is enhanced by the gating. This shifts the center of the electron distribution closer to the interface. For fixed sheet carrier density, $n_{2D}$, this increased confinement would shift $E_F$ upwards in energy, opposite to the effect of depletion. Since the SX-PES Ti 2$p$ core-level shifts to *higher* binding energy for $V_g$ < 0 V the band-bending induced upshift in $E_F$ is dominant. Indeed the dielectric constant of SrTiO$_3$ at room temperature $\varepsilon_r(T = 300$ K$) \sim 350$, gives a total carrier density change of $\sim 2 \times 10^{10}$ cm$^{-2}$ for $V_g$ = -50 V. This is just 0.14 % of the total Hall sheet carrier density. The magnitude of the energy shift clearly depends on the probing depth of the PES measurement compared to the characteristic length scale of the confinement potential narrowing. Hence the lack of an observed shift in the HAX-PES data suggests that the most significant changes in the confining potential occur in the topmost layers inside the SrTiO$_3$.

In order to quantitatively analyze these scenarios, we simulated the potential profile in SrTiO$_3$ as a function of the depth ($z$) from the interface, using the PES data to constrain the results. The scheme of self-consistent calculation is as follows: for a trial potential $\Phi(z)$, the carrier density profile $n(z)$ is calculated using

$$n(z) = \int_{\epsilon=0}^{\infty} D(\epsilon)f(\epsilon)d\epsilon, \tag{1}$$

where $D(\epsilon)$ is the density of states at electron energy $\epsilon$ from the conduction band minimum which is calculated via $D(\epsilon) = (2m^*/\hbar^2)^{3/2} \epsilon(z)^{1/2}/2\pi^2$, where $m^*$ is the effective mass of SrTiO$_3$, $\hbar$ is Plank's constant divided by $2\pi$, and $\epsilon(z)$ is based on $\Phi(z)$ [14], and $f(\epsilon)$ is the Fermi-Dirac function. The potential is obtained by solving Poisson's equation

$$\frac{\partial}{\partial z}\left(\varepsilon_0 \varepsilon_r \frac{\partial \Phi(z)}{\partial z}\right) = -e_0 n(z). \tag{2}$$

Here, we assumed other sources of charge, e.g. holes, extrinsic donors and acceptors, are absent. $\varepsilon_0$ is the vacuum permittivity, $\varepsilon_r$ is the bulk relative permittivity of SrTiO$_3$, and $e_0$ is the elemental charge. Equations (1) and (2) were solved self-consistently with a convergence criterion of 0.01 % throughout the whole depth region. For each $V_g$, the two boundary conditions used are $\Phi(z = 0)$ and $\frac{\partial \Phi(z)}{\partial z}|_{z=\infty}$. The former is equal to the energy shift of the Ti 2$p$ core-level peak position in the SX-PES spectra at the appropriate $V_g$ [25], and the latter is 0 V/m and $1\times10^5$ V/m, for $V_g = 0$ and -50 V respectively, given the 0.5 mm SrTiO$_3$ substrate thickness. The core-level spectrum $I(E)$ is computed according to

$$I(E) \propto \sum_{z=0}^{\infty} \exp(-z/\lambda \cos\theta) C[E - E_0 - \Phi(z)],$$

where $C(E)$ represents the core-level spectra described by a Voigt function peaked at $E_0$ [25]. The final best fit result is determined by minimizing the sum of the squares of the difference between the measured and calculated forms of the $I(E)$ for both the SX-PES and HAX-PES measurements.

The electric field ($\mathbb{E}$) dependent local nonlinear permittivity of SrTiO$_3$, $\varepsilon_\mathrm{r}(\mathbb{E}, z)$, must be considered as discussed in previous reports [14,22,28]. First we attempted to utilize the reported form of $\varepsilon_\mathrm{r}(\mathbb{E})$ in our simulation (Fig. 4(c) inset) [29]. However when $n_\mathrm{2D}$ was constrained to the measured value, satisfactory agreement with the PES spectra was not possible. Even allowing $n_\mathrm{2D}$ to be a free parameter (allowing for the possibility of localized charge not contributing to the Hall measurement), no value could reasonably fit the spectra and key features such as the core-level shift in Fig. 3(c) (see supplementary materials [30]). The essential issue is that the reported bulk form of $\varepsilon_\mathrm{r}(\mathbb{E})$ is not sufficiently nonlinear in electric field.

As an alternative, we then assumed a smoothly varying dielectric constant as a function of distance from the interface $\varepsilon_\mathrm{r}(z)$ for $V_\mathrm{g} = 0$ and -50 V, based on a simple sigmoid function. The self-consistently solved potential and carrier density profiles are shown in Figs. 4(a) and (b). Here the errors of the simulated parameters were estimated with a threshold of 3 % increase in the total squared error from the minimum value, except for the error in $\Phi(0)$ which was directly taken from the experimental error in the PES spectra. The comparison between the experimental spectra (open symbols) and the best-fit simulated spectra (solid lines) from the obtained potential profiles are shown in Figs. 4(d) and (e). Good agreement is observed for both the SX-PES and HAX-PES data. The main panel of Fig. 4(c) shows the best fit $\varepsilon_\mathrm{r}(z)$ for $V_\mathrm{g} = 0$ and -50 V in these simulations, and the insets of Fig. 4(c) shows the extracted $\varepsilon_\mathrm{r}(\mathbb{E})$.

From these simulations, close to the interface, we find a relatively abrupt potential shift of ~0.15 eV inside the SrTiO$_3$ for $V_g$ = -50 V compared to $V_g$ = 0 V. As shown in the inset of Fig. 4(a), the potential crosses $E_F$ around $z$ = 6 nm and 2 nm for $V_g$ = 0 and -50 V, the former in good agreement with previous experimental and theoretical estimates of the electron gas thickness being < 10 nm [11,13,14,31].

Experimentally a significant change of the potential around the interface occurred only for $V_g$ < 0 V, and not $V_g$ > 0 V. Based on these calculation results, we can explain this asymmetry by considering the magnitude of $\mathbb{E}$ around $z$ = 0. For $V_g$ = 0 V, $\mathbb{E}$ is not large enough to significantly reduce $\varepsilon_r$, and positive $V_g$ only tends to decrease the interfacial electric field. The resultant changes in the band-bending and position of $E_F$ are therefore relatively small and below the PES resolution. This is in stark contrast to the case of $V_g$ < 0 V, where $\mathbb{E}(z = 0)$ is large enough to reduce $\varepsilon_r$, which self-consistently enhances the confinement leading to significant changes in the potential which are measurable by SX-PES.

Quantitatively, the collapse of $\varepsilon_r$ with $\mathbb{E}$ obtained in the analysis above is more rapid than reported for non-doped SrTiO$_3$ [14,32-34] and Nb doped SrTiO$_3$ [29,35]; the latter being somewhat faster than the former one. This has important implications for calculations of electron accumulation layers in any SrTiO$_3$–based heterostructure. We note that our self-consistent approach is the same as Copie *et al.*, who utilized the literature form of $\varepsilon_r(\mathbb{E})$ [14]. The $\varepsilon_r(\mathbb{E})$ relationship reported by Yamamoto *et al.* [29,35] has also been successfully used to model the depletion layer in metal/SrTiO$_3$ Schottky junctions [29,36,37]. An important difference between the Schottky depletion layer and the LaAlO$_3$/SrTiO$_3$ is the existence of free electrons in the latter, which can screen applied electric fields in addition to the lattice polarization which is the only possibility in the former. Indeed, a recent theoretical study has noted the interplay between electron density changes and lattice polarization [38]. In order to clarify these points, direct microscopic investigations of the lattice polarization with gate voltage are essential. Noting the large changes in potential over just a few lattice parameters, the failure of the prior experimental measurements of $\varepsilon_r(\mathbb{E})$ to capture our data likely reflect the need to explicitly consider $\varepsilon_r$ on short length scales, and include nonlocal effects [39,40].

Finally we note the intriguing point that when the electric displacement field $D(\mathbb{E})$ is calculated using the new functional form of $\varepsilon_r(\mathbb{E})$, $\mathbb{E}$ is multi-valued for $0.02 \leq D \leq 0.08$ Cm$^{-2}$, as shown in Fig. 4(f). Although the underlying physics and effects of such an electrostatic instability are currently not clear, it is possible that the presence of multiple metastable dielectric states in SrTiO$_3$ close to the interface could induce local structural phase transitions [41] and associated effects in resistive switching properties [42], in addition to creating an unstable potential profile at the interface. The collapse of $\varepsilon_r$ around the interface, simultaneously enhancing the electron confinement and impurity scattering, especially at low temperatures, explains the substantial decrease in the mobility [22] and enhanced localization [41] that has been previously observed for back-gating. The strong contrast with top-gating [43] suggests that this nonlinear dielectric response provides new device switching approaches in oxide heterostructures.

The authors are grateful to R. Yasuhara, and S. Toyoda for experimental support, M. S. Bahramy, R. Takahashi, and M. Lippmaa for the useful discussions. This work was primarily supported by the Department of Energy, Office of Basic Energy Sciences, Division of Materials Sciences and Engineering, under contract DE-AC02-76SF00515 (M.M., Y.H., C.B., H.I., M.H., H.K.S. and H.Y.H.). M.M. acknowledges partial support from the ONR-MURI number N00014-12-1-0976.

**References**

1) A. Ohtomo and H. Y. Hwang, Nature **427**, 423 (2004).

2) N. Reyren *et al.*, Science **317**, 1196 (2007).

3) A. D. Caviglia *et al.*, Nature **456**, 624 (2008).

4) A. Brinkman *et al.*, Nat. Mater. **6**, 493 (2007).

5) D. A. Dikin *et al.*, Phys. Rev. Lett. **107**, 056802 (2011).

6) Ariando *et al.*, Nat. Commun. **2**, 188 (2011).

7) D. G. Schlom, and J. Mannhart, Nat. Mater. **10**, 168 (2011).

8) N. Nakagawa, H. Y. Hwang, and D. A. Muller, Nat. Mater. **5**, 204 (2006).

9) P. R. Willmott *et al.*, Phys. Rev. Lett. **99**, 155502 (2007).

10) A. Kalabukhov *et al.*, Phys. Rev. B **75**, 121404 (2007).

11) W. Siemons *et al.*, Phys. Rev. Lett. **98**, 196802 (2007).

12) G. Herranz *et al.*, Phys. Rev. Lett. **98**, 216803 (2007).

13) M. Basletic *et al.*, Nat. Mater. **7**, 621 (2008).

14) O. Copie *et al.*, Phys. Rev. Lett. **102**, 216804 (2009).

15) K. Yoshimatsu *et al.*, Phys. Rev. Lett. **101**, 026802 (2008).

16) M. Sing *et al.*, Phys. Rev. Lett. **102**, 176805 (2009).

17) M. Takizawa *et al.*, Phys. Rev. B **84**, 245124 (2011).

18) G. Berner *et al.*, Phys. Rev. B **88**, 115111 (2013).

19) S. Thiel *et al.*, Science **313**, 1942 (2006).

20) C. Cen *et al*., Nat. Mater. **7**, 298 (2008).

21) A. Joshua *et al*., Nat. Commun. **3**, 1129 (2012).

22) C. Bell *et al.*, Phys. Rev. Lett. **103**, 226802 (2009).

23) D. Rakhmilevitch *et al*., Phys. Rev. B **87**, 125409 (2013).

24) N. Reyren *et al*., Phys. Rev. Lett. **108**, 186802 (2012).

25) M. Minohara *et al*., Phys. Rev. B **85**, 165108 (2012).

26) N. Ohashi *et al*., Appl. Phys. Lett. **101**, 251911 (2012).

27) S. Tanuma, C. J. Powell, and D. R. Penn, Surf. Interface Anal. **21**, 165 (1994).

28) J. Biscaras *et al*., Phys. Rev. Lett. **108**, 247004 (2012).

29) T. Yamamoto *et al.*, Jpn. J. Appl. Phys. **36**, L390 (1997).

30) See Supplemental Material for more details on calculations.

31) S. Su, J. H. You, and C. Lee, J. Appl. Phys. **113**, 093709 (2013).

32) R. C. Neville, B. Hoeneisen, and C. A. Mead, J. Appl. Phys. **43**, 2124 (1972).

33) C. Ang *et al*., J. Appl. Phys. **87**, 3937 (2000).

34) H.-M. Christen *et al*., Phys. Rev. B **49**, 12095 (1994).

35) T. Yamamoto *et al*., Jpn. J. Appl. Phys. **37**, 4737 (1998).

36) Y. Hikita *et al*., Appl. Phys. Lett. **90**, 143507 (2007).

37) T. Susaki *et al*., Phys. Rev. B **76**, 155110 (2007).

38) G. Khalsa, and A. H. Macdonald, Phys. Rev. B **86**, 125121 (2012).

39) M. Stengel and N. A. Spaldin, Nature **443**, 679 (2006).

40) D. R. Hamann, D. A. Muller, and H. Y. Hwang, Phys. Rev. B **73**, 195403 (2006).

41) M. Rössle *et al*., Phys. Rev. Lett. **110**, 136805 (2013).

42) S. Wu *et al*., Phys. Rev. X **3**, 041027 (2013).

43) M. Hosoda *et al*., Appl. Phys. Lett. **103**, 103507 (2013).

**Figure captions:**

FIG. 1 (color online). Experimental setup for measurements of PES spectra for LaAlO$_3$/SrTiO$_3$ heterostructures with applied electric field at the back of the SrTiO$_3$. Ground contact is made to the electron gas at the heterointerface.

FIG. 2 (color online). Ti 2$p$ core-level spectra of LaAlO$_3$/SrTiO$_3$ heterostructures without bias gate voltage. Solid lines and dashed lines correspond to 4 nm and 2 nm LaAlO$_3$ samples, respectively. The relative binding energy is given with respect to the Ti 2$p$ core-level spectra of a bare SrTiO$_3$ substrate as reference. $\lambda$ is the probing depth controlled by photon energy and emission angle.

FIG. 3 (color online). Ti 2$p$ core-level spectra of LaAlO$_3$/SrTiO$_3$ heterostructures measured using (a) SX-PES and (b) HAX-PES with applied gate voltage. Plots of relative binding energy (B.E.) shift between (c) Ti 2$p$ and La 4$d$ core-level spectra (filled circles), Ti 2$p$ and Al 2$p$ (open triangles) for SX-PES data, and (d) Ti 2$p$ and La 4$d$ core level spectra (filled circles) from HAX-PES data.

FIG. 4 (color online). (a) Simulated potential depth profiles of the electron gas, for $V_g = 0$ and -50 V. Inset of Fig. 4(a) shows the magnification around the interface. (b) Self-consistent carrier profile and (c) the resultant $\varepsilon_r(z)$ for $V_g = 0$ and -50 V. The inset of Fig. 4(c) shows the comparison between the extracted form of $\varepsilon_r(\mathbb{E})$ (symbols and dashed line) and the reported $\varepsilon_r(\mathbb{E})$ (solid line) [29]. Measured and simulated SX-PES and HAX-PES Ti 2p core-level spectra for (d) $V_g = 0$ V, and (e) $V_g = -50$ V. Open circles are the experimental data, and solid lines are the best-fit simulations. Dashed lines correspond to the Ti 2p core-level spectrum of a bare SrTiO$_3$ substrate. (f) The calculated electric displacement field from the inset of Fig. 4(c). Solid line is calculated by fitting the extracted form of $\varepsilon_r(\mathbb{E})$ in the inset of Fig. 4(c) using a sigmoid function.

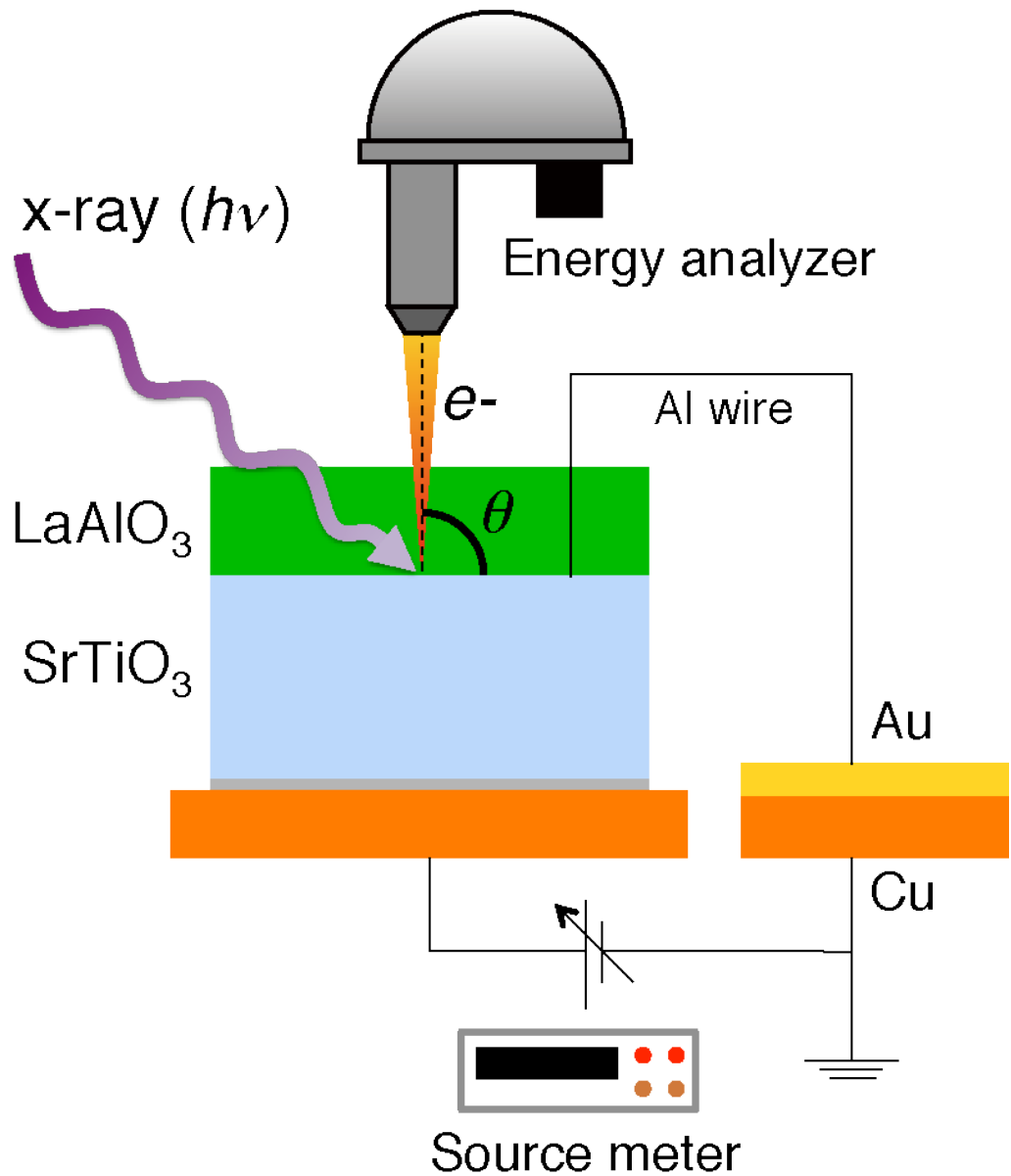

Figure 1. M. Minohara *et al*.

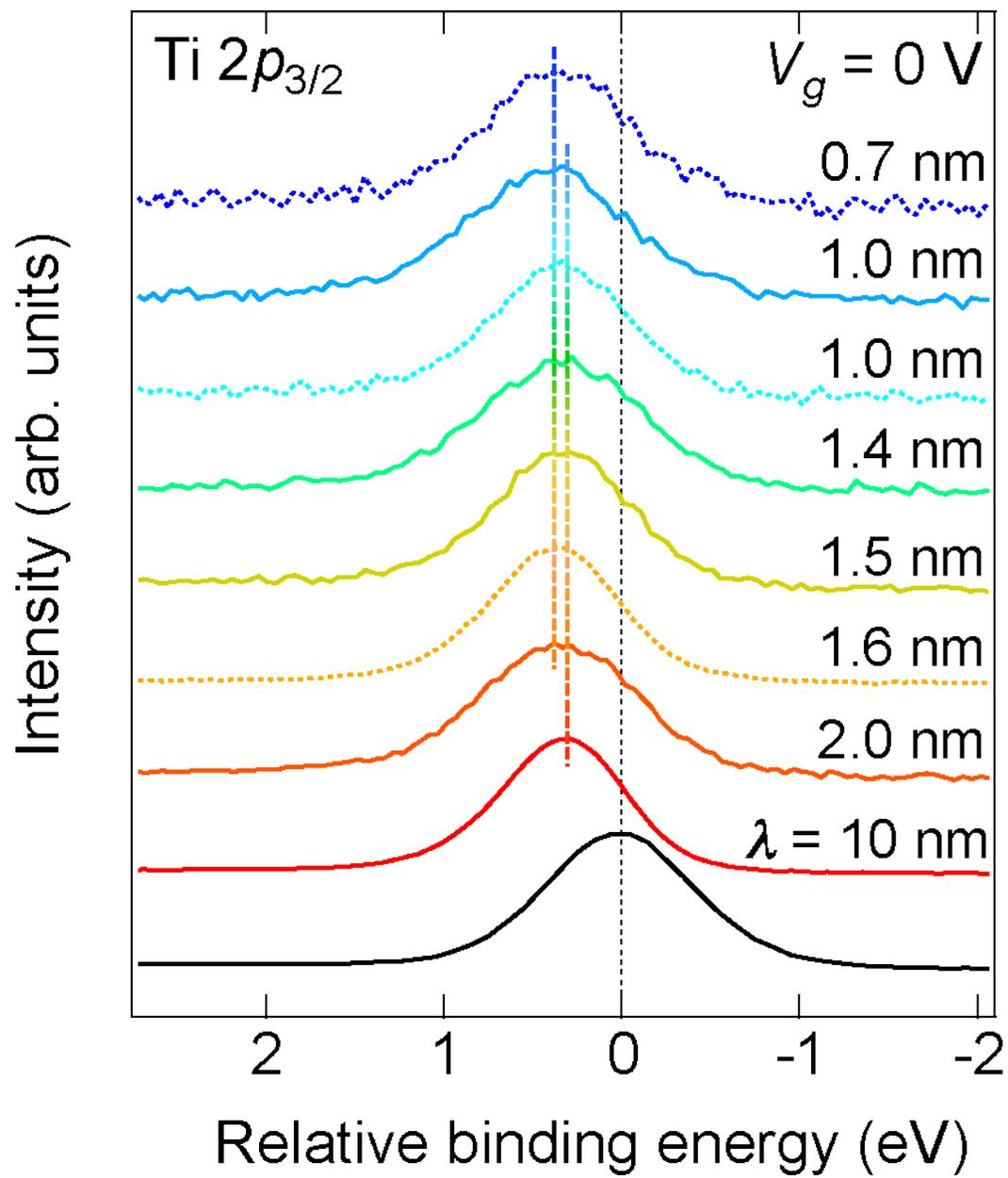

Figure 2. M. Minohara *et al*.

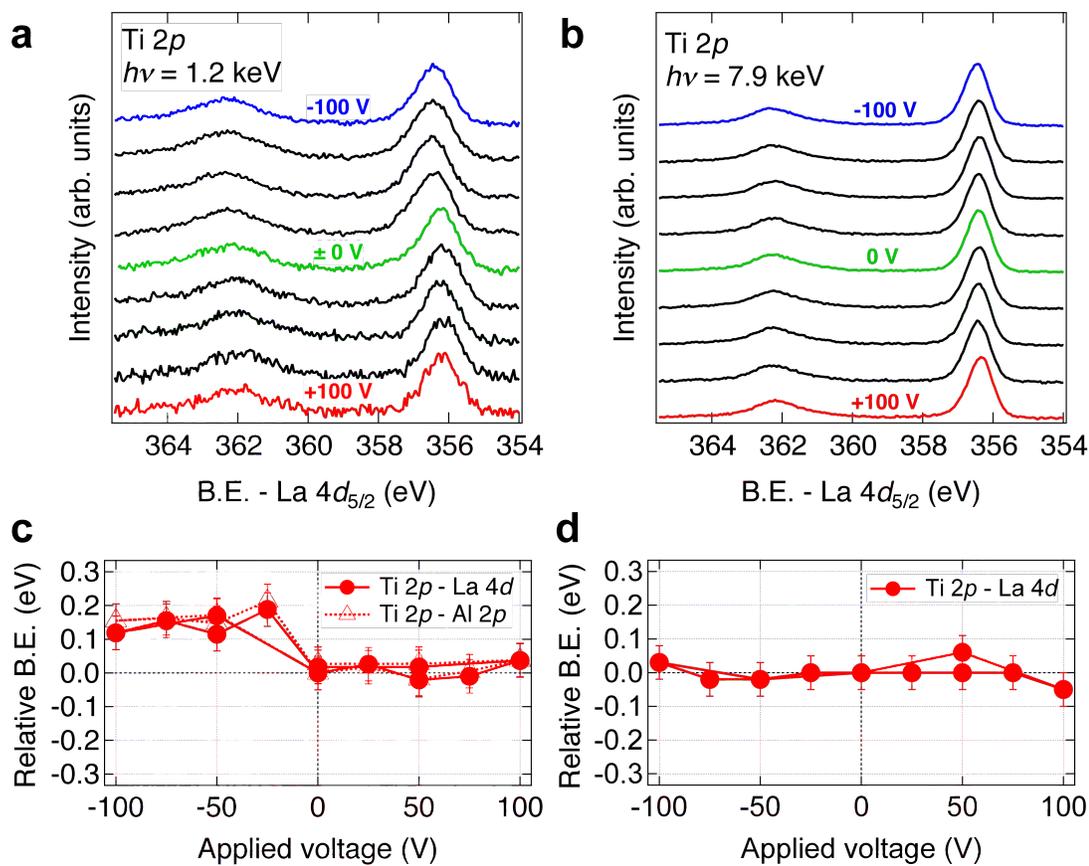

Figure 3.  M. Minohara *et al*.

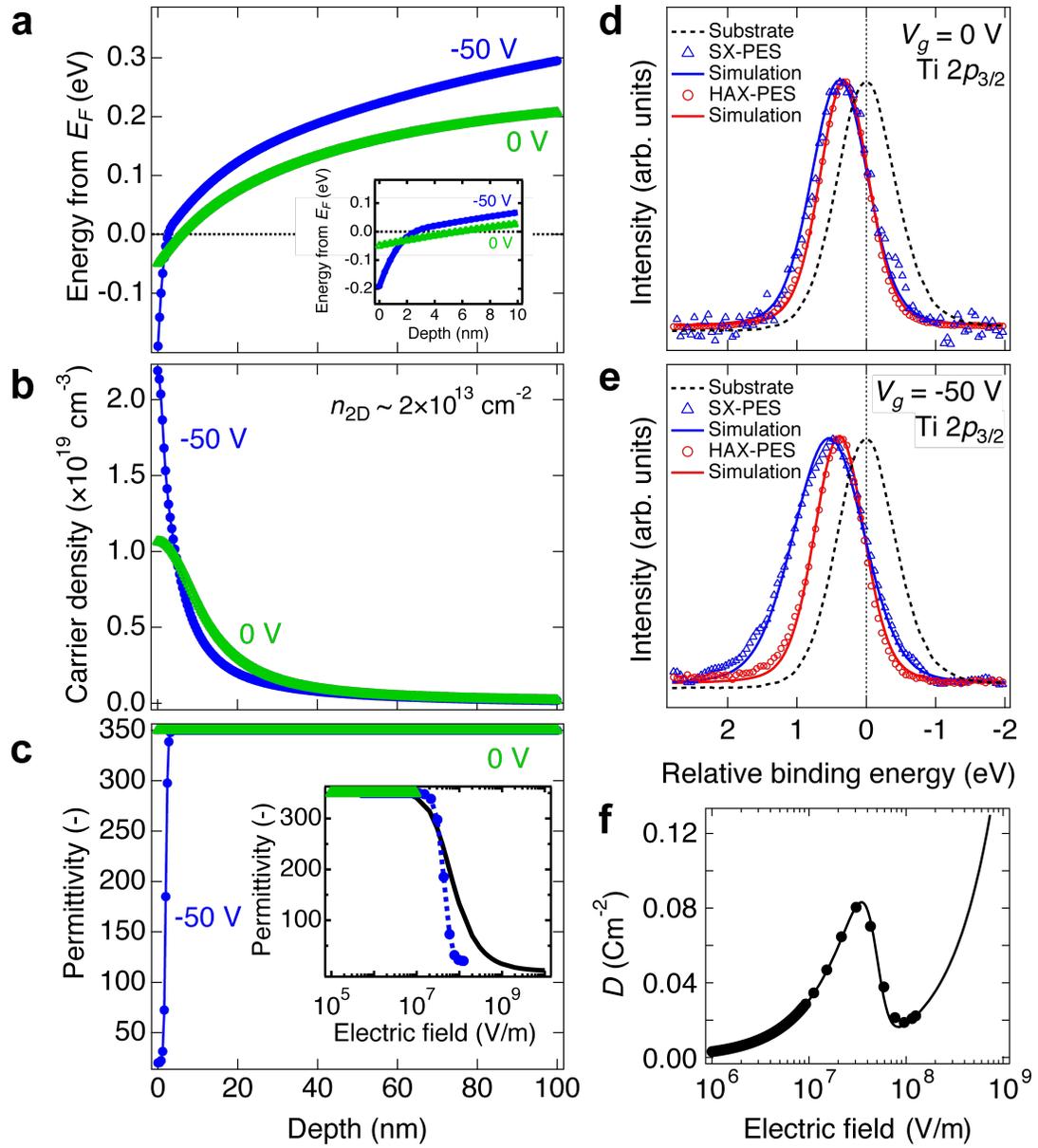

Figure 4. M. Minohara *et al*.

# The potential profile at the LaAlO$_3$/SrTiO$_3$ (001) heterointerface in operando conditions

M. Minohara, Y. Hikita, C. Bell, H. Inoue, M. Hosoda, H. K. Sato, H. Kumigashira, M. Oshima, E. Ikenaga, and H. Y. Hwang

Author to whom correspondence should be addressed; Electronic mail: minohara@stanford.edu

**This PDF file includes:**

Fig. S1

**Potential profiling simulation based on reported nonlinear permittivity**

In order to achieve a full quantitative simulation of the gating dependence of the interface, the local nonlinear permittivity $\varepsilon_r(\mathbb{E}, z)$ of SrTiO$_3$ must be considered. Therefore, we initially tried to utilize the reported $\varepsilon_r(\mathbb{E})$ in our simulation [S1]. Figure S1 shows the potential profiling analysis for LaAlO$_3$/SrTiO$_3$ corresponding to the measured sheet carrier density of $n_{2D} \sim 2\times10^{13}$ cm$^{-2}$ [S2].

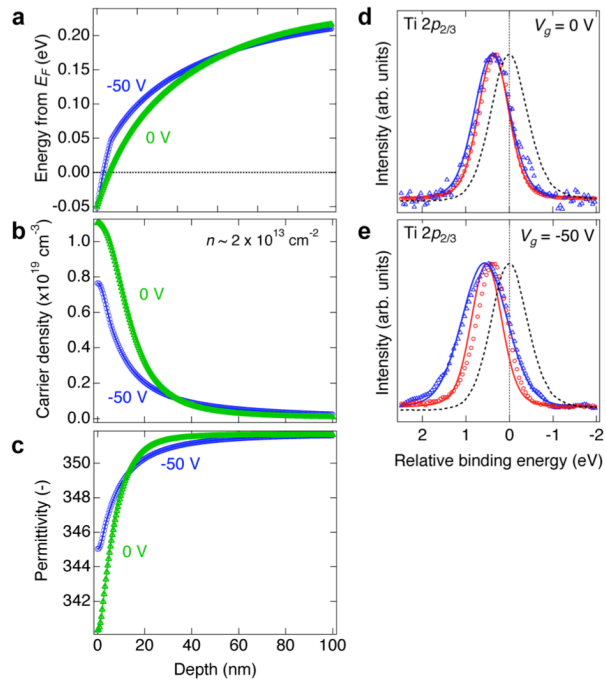

Fig. S1 Self-consistently solved (a) potential profile, (b) carrier profile, and (c) permittivity profile in depth for $V_g = 0$ (green) and -50 V (blue) corresponding to the measured sheet carrier density. The Ti 2$p$ core-level spectra of buried SrTiO$_3$ substrate obtained by SX-PES and HAX-PES measurements (blue and red, respectively) under (d) $V_g = 0$ V, and (e) -50 V, respectively. Open circles are the experimental results, and solid lines are the simulated results. Black dashed line corresponds to the Ti 2$p$ core-level spectra of bare SrTiO$_3$ substrate.

In this condition, the potential profile cannot explain the measured core-level spectra for $V_g$ = -50 V as shown by the poor quality of the fit in Fig. S1(e). In addition, Fig. S1(a) does not show any shift at the edge of the potential, indicating there should be no core-level shifts, while we obviously observed the core-level shift by applying negative gate voltage as shown in Fig. 3. Allowing $n_{2D}$ to vary did not resolve these issues, which motivated the use of a sigmoid function for $\varepsilon_r(z)$ as discussed in the main text.

**References**

S1. T. Yamamoto *et al.*, Jpn. J. Appl. Phys. **36**, L390 (1997).

S2. C. Bell *et al.*, Phys. Rev. Lett. **103**, 226802 (2009).